\documentstyle[12pt]{article}
\newcommand{\be}{\begin{equation}}
\newcommand{\ee}{\end{equation}}
\newcommand{\bea}{\begin{eqnarray}}
\newcommand{\eea}{\end{eqnarray}}
\newcommand{\la}{\label}
\newcommand{\noi}{\noindent}
\newcommand{\nn}{\nonumber}

\newcommand{\al}{\alpha}
\newcommand{\Si}{\Sigma}
\newcommand{\lm}{\lambda}
\newcommand{\g}{\gamma}

\newcommand{\dl}{\delta}
\newcommand{\ep}{\varepsilon}
\newcommand{\e}{\epsilon}

\newcommand{\tr}{\mathop{{\rm tr}}}
\newcommand{\trp}{\mathop{\mbox{tr P}}}

\newcommand{\dxm}{\mathop{dx^{\mu}}}
\newcommand{\ddx}{\mathop{dx^{\mu}\wedge dx^{\nu}}}
\newcommand{\dt}{\mathop{dt}}
\newcommand{\dlm}{\int_{1-\varepsilon}^{1+\varepsilon}
\mathop{d\lambda}}
\newcommand{\de}{\int\limits_{-\infty}^{\infty}
\mathop{\frac{d\eta}{2\pi}}}
\newcommand{\Dzi}{\int_{PBC} \mathop{D^2z} \delta(\int^1_0
\mathop{dt} z^{\dag}z-1-\frac{N}{2})}
\newcommand{\intt}{\int^1_0}
\newcommand{\wl}{\mathop{\mbox{tr
P}}e^{i\oint_{\gamma}\mathop{dx^{\mu}} A_{\mu}(x)}}

\newcommand{\dT}{-\frac 1 2 \int_0^{\infty} \frac{dT}{T} \:
e^{-m^2T}}
\newcommand{\dTM}{-\frac 1 2 \int_0^{\infty} \frac{dT}{T} \:
e^{-im^2T}}
\newcommand{\dQ}{\int_{PBC} DqD^2\psi D^2 z\delta\left( z^{\dag}z
-1-\frac N 2 \right)\delta\left( \psi^{\dag}\psi -3\right)}
\newcommand{\dQM}{\int_{PBC} DqD^2\psi D^2 z\delta\left( z^{\dag}z
-1-\frac N 2 \right)\delta\left( \bar{\psi}\psi -3\right)}
\newcommand{\wH}{(-\nabla_{\mu}\nabla^{\mu}+
\sigma^{\mu\nu}F_{\mu\nu})}

\newcommand{\limd}{\lim_{\delta \to +0}}
\newcommand{\lime}{\lim_{\epsilon \to +0}}

\newcommand{\zd}{z^{\dag}}
\newcommand{\nh}{\hat{\nabla}}
\newcommand{\pdd}{\psi^{\dag}}
\newcommand{\xd}{\xi^{\dag}}
\newcommand{\co}{coset $SU(N)/U(1)\times SU(N-1)${}}
\newcommand{\pbb}{\bar{\psi}}

\newcommand{\oeo}{
\left( -\frac d{dt}e^{\e\frac d{dt}} + iB\right)}
\newcommand{\oee}{
\left( -\frac d{dt} + iBe^{-\e\frac d{dt}}\right)}
\newcommand{\owe}{
\left( -\frac d{dt} + iB\right)}

\title{Pure bosonic worldline path integral \\ representation for
fermionic determinants, non-Abelian Stokes theorem, \\ and
quasiclassical approximation in QCD}

\author{F. A. Lunev \thanks{electronic address:
lunev@hep.phys.msu.su } \\ {\em Physical Department, Moscow State
University,} \\ {\em Moscow, 119899, Russia} }

\date{ \ \ \  }

\begin{document}

\maketitle

\begin{abstract}
Simple bosonic path integral representation for path ordered
exponent is derived. This representation is used, at first, to
obtain new variant of non-Abelian Stokes theorem. Then new pure
bosonic worldline path integral representations for fermionic
determinant and Green functions are presented. Finally, applying
stationary phase method, we get quasiclassical equations of motion
in QCD.
\end{abstract}

\section{Introduction}

Elimination of fermionic degrees of freedom is very desirable in
many problems of quantum field theory and elementary particle
physics. First, in lattice gauge theories integration with respect
to Grassmannian variables leads to serious complications of
numerical simulations. The second, appearance of fermionic variables
in functional integrals hampers the application of stationary phase
method. As a consequence, one cannot also apply quasiclassical
expansions to evaluation of functional integrals in theories with
fermions except some especially simple cases.

Indeed, typical Green function can be written as

\begin{equation}
G(x_1, ..., x_n) = \int DA D\Psi D\bar{\Psi} e^{\frac{i}{\hbar}
(S_{YM}(A) + S_{ferm}(\bar{\Psi}, \Psi, A))}
{\cal O}_1(x_1) ...{\cal O}_n(x_n)
\label{1}
\end{equation}

\noindent where ${\cal O}_i$ are some operators whereas $S_{YM}$ and
$S_{ferm}$ are Yang-Mills and fermionic actions respectively.
Quasiclassical approximation is defined, up to some subtleties, by
stationary point equations

\begin{equation}
\frac{\delta S_{YM}}{\delta A}=[\mbox{the source of YM field}]
\label{2}
\end{equation}

But what must be written in R.H.S. of eq. (\ref 2) ? Of course, one
cannot put

\begin{equation}
[\mbox{the source of YM field}]=\frac{\delta S_{ferm}}{\delta A}
\label 3
\end{equation}

\noindent because $S_{ferm}$ depends on Grassmanian variables.
Moreover, without preliminary exception of fermionic variables one
cannot write in R.H.S. of (\ref 2) nothing except zero. But this
means that in zero approximation YM field can be considered as free.
It seems inappropriate in all cases in which interaction is strong.

So for application of quasiclassical methods as well as for
facilitation of numerical simulations on the lattice fermionic
variables in functional integrals of the type (\ref 1)
must be integrated out and result must be represented as functional
integral with respect to only bosonic variables. In other words, the
theory must be bosonized.

The problem of bosonization of fermionic theories has a long
history. Most likely, the first example of bosonisation of fermionic
theory was given by Schwinger in his famous paper \cite{1}
concerning full solution of massless  $\mbox{QED}_2$. Then the problem
of bosonisation was investigated by many authors, but more or less
exhaustive solution was obtained only in two dimensional case and
in some three dimensional models (see, for instance, papers \cite{2}
and references therein).  In realistic four dimensional case only
partial success was achieved (see, for instance, \cite 3). In fact in
all proposed bosonization schemes in four dimensions  it is necessary
to evaluate (exactly or in some approximation) fermionic determinant
-- but it is just the main problem that must be solved by means of
bosonization.

To author's knowledge, the only exceptions are recent papers by
Lusher \cite 4 and Slavnov \cite 5 (see also \cite 6).  In Ref. \cite
4 fermionic determinant on the finite lattice is represented as
infinite some of bosonic determinants. In Ref. \cite 5  fermionic
determinant in $D$ dimensions is expressed via bosonic one in $D+1$
dimensions. These approaches seem useful in lattice theories
but they cannot be applied to investigation of
quasiclassical approximation.

So hitherto no quite satisfactory representation for fermionic
determinant in terms of bosonic fields is known, and at present
paper we will develop another approach to bosonization. Namely, we
will derive pure bosonic worldline path integral representation for
fermionic determinants, Green functions and Wilson loops.

Worldline approach to quantum field theory also has very long
history. It was originated many years ago in classical works by
Feynman \cite 7 and Schwinger \cite 8 . The main idea of this
approach is to represent fermionic determinants and fermionic Green
functions as functional integral over trajectory of a single
relativistic particle.

Let us consider, for instance, fermionic determinant for $SU(N)$
Yang-Mills theory in Euclidean space:

\begin{equation}
D \equiv \det (i\hat{\nabla} +im)
=\det (i\hat{\nabla} +im)\gamma^5
\label 4
\end{equation}

\noindent where $\hat{\nabla}=\gamma^{\mu} \nabla_{\mu} =\gamma^{\mu}
(\partial_{\mu} - iA_{\mu}), \ \ (\gamma^{\mu})^{\dag}=\gamma^{\mu},
\ \ \{\gamma^{\mu},\gamma^{\nu}\}=2\delta^{\mu \nu}, \ \
iA_{\mu}(x)\in su(N)$.

One can write:
\begin{equation}
\ln D =\frac{1}{2} \ln \det[(i\hat{\nabla} +im)\gamma^5]^2=
\frac{1}{2} \ln {\det} (-\nabla_{\mu}\nabla_{\mu} +
\sigma^{\mu\nu}F_{\mu\nu} +m^2)
\label 5
\end{equation}

\noindent where

\begin{equation}
F_{\mu\nu}=\partial_{\mu}A_{\nu}-\partial_{\nu}A_{\mu}-
i[A_{\mu},A_{\nu}],
\label 6
\end{equation}

\begin{equation}
\sigma^{\mu\nu}=\frac{i}{4}[\gamma^{\mu},\gamma^{\nu}]
\label 7
\end{equation}

Further, eq. (\ref 5) can be written, up to inessential constant, as

\bea
\ln D&=&
\frac{1}{2} \mbox{tr}\: \ln (-\nabla_{\mu}\nabla_{\mu} +
\sigma^{\mu\nu}F_{\mu\nu} +m^2)\nn\\
&=&-\frac{1}{2} \int_0^{\infty}\frac{dT}{T} e^{-m^2T}
 \mbox{tr}\:
e^{-T(
-\nabla_{\mu}\nabla_{\mu} +
\sigma^{\mu\nu}F_{\mu\nu})}
\label 8
\eea

The integral (\ref 8) is divergent at $T=0$ and must be regularized.
One can use, for instance, $\zeta$-function regularization:

\begin{equation}
\ln D =-\frac{1}{2} \frac{d}{ds} \frac{1}{\Gamma(s)} \int_0^{\infty}
dT \ T^{s-1} e^{-m^2 T}
 \mbox{tr}\:
e^{-T(
-\nabla_{\mu}\nabla_{\mu} +
\sigma^{\mu\nu}F_{\mu\nu})}
\raisebox{-9pt}{\rule{0.4pt}{20pt}${\scriptstyle s=0}$}
\label 9
\end{equation}

\noi However, in what follows we shall write, for short,         formal
expression (\ref 8) for $\ln D$ bearing in mind any suitable
regularization.

Further, trace in eq.(\ref 8) can be represented as functional
integral:

\bea
&& \mbox{Tr}
e^{-T(
-\nabla_{\mu}\nabla_{\mu} +
\sigma^{\mu\nu}F_{\mu\nu})}\nn\\
&=& \int _{PBC} Dq\: \mbox{tr P}\exp \left\{ -\int_0^1 dt
\left(\frac{\dot{q}^2}{4T} -i\dot{q}^{\mu}A_{\mu}(q) +T
\sigma^{\mu\nu}F_{\mu\nu}(q) \right) \right\}
\label{10}
\eea

\noindent Here PBC means 'periodic boundary conditions'. Path
ordering in (\ref{10}) corresponds to both spin and colour matrix
structures.

Let us suppose for a moment that spin and colour degrees of
freedom are absent in (\ref{10}). (This is just the case of scalar
electrodynamics). Then, integrating in (\ref 1) over fermionic
fields and using for arising fermionic determinants and fermionic
Green functions formulae of the type (\ref 8), (\ref{10}), one
obtains formulation of quantum field theory in terms of particles
interacting with gauge field $A$. This is just result of classical
work bu Feynman \cite 7.

In realistic models, however, one must take into account spin and
colour degrees of freedom. So, to develop worldline formulation of
quantum field theory, it is necessary to represent the path ordered
exponent in (\ref{10}) as functional integral. In the case of QED
this was done by Fradkin \cite 9 in terms of fermionic path
integral. This representation and their modifications were
successfully used, in particular, for construction of derivative
expansion in QED \cite{10} and for investigation of complicated
Feynman diagrams \cite{11}. It is also intensively used for
investigation of hidden supersymmetry in fermionic theories (see,
for example, \cite{12} and references therein). Recently, D'Hoker
and Gagne derived fermionic path integral representation for
fermionic determinants for particles coupled with arbitrary tensor
field \cite{13}.

In papers \cite{8}-\cite{13} fermionic path integral representations
were derived and used only for spin degrees of freedom. For colour
P-exponent in (\ref{10}) fermionic path integral representation also
can be derived (see, for example, \cite{14}) though it seems not so
elegant as one for spin P-exponent. Apropos, works \cite{14} were, to
author's knowledge, the first attempts to obtain non-perturbative
information about QCD in worldline formalism.

But, as we already stated at the beginning of the present paper,
there exist important problems for which pure bosonic worldline path
integral representation for fermionic determinants and Green
functions is very desirable. We see that for solution of the latter
problem it is sufficient to derive bosonic worldline path integral
representation for the trace of path ordered exponent

\begin{equation}
Z=\mbox{tr P} e^{i\int_0^1 dt B(t)}
\label{11}
\end{equation}

\noindent with matrix $B(t) \in GL(N),\;  B(0)=B(1)$. Indeed,
substituting such representation with

\begin{equation}
B(t)=\dot{q}^{\mu}A_{\mu}(q(t)) - \sigma^{\mu\nu}F_{\mu\nu}(q(t))
\label{12}
\end{equation}

\noindent in (\ref{10}), one obtains desired representation for
fermionic determinant. As we will show later, bosonic path integral
representation for $Z$ allows also to obtain bosonic worldline path
integral representation for fermionic Green functions.

Different bosonic path integral representations for $Z$ in the case
$B(t) \in SU(2)$ were proposed by several authors (see
\cite{15}-\cite{18}). For more general case $B(t) \in su(N)$ the an
analogous representation was pointed out in \cite{19}.

The typical result for $B \in SU(2)$ is

\begin{equation}
Z=\int DS(t) \exp \left\{ \frac{i}{2} \int _0^1 dt \left[ \mbox{tr}
\sigma^3 \left( S(t)B(t)S^{\dag}(t) +iS(t)\dot{S}^{\dag}(t) \right)
\right] \right\}
\label{13}
\end{equation}

The integration in (\ref{13}) is carried out over all trajectories
in the group $SU(2), \; S \in SU(2), \; \sigma^3$ is a Pauli matrix.

The last result has two disadvantages. First, in any parametrization
of $SU(2)$ the "action" in (\ref{13}) is rather complicated
non-polinomial function. This hampers the usage of (\ref{13}) in
practice. The  second, it appears that integral (\ref{13}) is
 ill-defined and needs insertion of some
 regulators in the "action" (see
the discussion of this point in \cite{16,19} and in more recent
paper \cite{20}).

Recently Dyakonov and Petrov found much more elegant formula for
path ordered exponent. Namely, for the case when $Z$ is Wilson loop
and gauge group is $SU(2)$, they derived from (\ref{13}) the
following expression for $Z$:

\bea
Z&=&\trp
e^{ i\oint_{\gamma} \dxm A_{\mu}(x) }\nn\\
&=&\int \mathop{Dn(x)} \prod_{x\in\Si} \d(n^a(x) n^a(x) -1)
\nn\\
&&\exp\left\{
\frac{i}{4}\int_{\Si}\ddx \left[ -F^a_{\mu\nu}n^a +\ep^{abc}n^a
D_{\mu}n^b D_{\nu} n^c \right] \right\}
\la{14}
\eea

\noi where $\Si$ is two dimensional surface spanned on contour
$\g$, $a=1,2,3$, $D^{\mu}$ -- covariant derivative. This formula can
be considered as non-Abelian variant of Stokes theorem. We shall
continue  the discussion of this result in section~4.

In the present paper we will derive alternative bosonic path
integral representation for $Z$ in the general case $B\in su(N)$.
The "action" in this representation is quadratic and so it is much
more simpler then one in (\ref{13}) and (\ref{14}). This derivation
is presented in the section 2. In the section 3 we check the
representation obtained by direct evaluation of the functional
integral that defines $Z$. In fact, we give alternative proof of
results obtained in the section 2. In the section 4 we  derive
an non-Abelian analog of Stokes theorem and compare our results with
those due to Dyakonov and Petrov. In the section 5 we derive bosonic
worldline path integral representation for fermionic determinant and
Green functions in Euclidean QCD. In the section 6 we get analogous
results for Minkowski space and then, applying stationary phase
method, derive quasiclassical equations of motion in QCD. In the
last section we summarize our results and discuss perspectives of
future investigations. In two appendixes we derive some auxiliary
formulae that are used in the main text of the paper.

\section{Bosonic path integral representation for the trace of path
ordered exponent}

Let $N\times N$ matrix $U(t)$, $0\le t\le 1$, be defined by equations

\bea
\frac{dU}{dt}&=&iB(t)U(t) \nn\\
U(0)&=&I_N
\la{15}
\eea

\noi where $I_N$ is $N\times N$ unit matrix. Then, obviously,

\be
Z=\tr U(1)
\la{16}
\ee

First, we consider the case $B \in su(N)$, that is

\be
B^{\dag }=B, \ \ \tr B=0
\la{16.1}
\ee

Let us consider an operator

\be
\hat B = a^{\dag}_r B^r_s a^s
\la{17}
\ee

\noi that acts in some Fock space ${\cal F}$. Operators $a^{\dag}_r$
 and $a^s$ in (\ref{17}) are usual {\it bosonic} creation and
annihilation ones, $[a^r,a^{\dag}_s]=\dl^r_s$.

Let ${\cal H}_n$ be $n$-particle subspace of ${\cal F}$, $\Pi_n$ is
 orthogonal projector on       ${\cal H}_n$. Then

\be
 Z= \tr \Pi_1 \mathop{{\rm P}} e^{i\int_0^1 \dt \hat B (t)}
\la{18}
\ee

Indeed, if $\hat N=a^{\dag}_r a^r$, then

\be
\left[ \hat B, \hat N \right] =0
\la{19}
\ee

So $\hat B {\cal H}_n \subset {\cal H}_n$ and in one dimensional
subspace ${\cal H}_1$ the operator $\hat B$ can be identified with
the matrix $B$ via relation:

\be
\hat B a^{\dag}_r \phi^r |0> = a^{\dag}_r (B^r_s\phi^s) |0>
\la{20}
\ee

Projector $\Pi_1$ can be represented as

\bea
\Pi_1&=&\dlm \dl(\hat N-\lm)\nn \\
&=&\dlm\de e^{-i\lm\eta + i\hat N \eta}
\la{21}
\eea

\noi where $0<\ep<1$.

So

\bea
Z&=&\limd \tr \Pi_1 e^{-\dl\hat N} \mathop{{\rm P}}
 e^{i\int_0^1 \dt \hat B (t)} \nn \\
&=&\limd \dlm \de \tr e^{-i\lm\eta +(i\eta-\dl)\hat N}
 e^{i\int_0^1 \dt \hat B (t)}\nn\\
&=&\limd \dlm \de \tr e^{-i\lm\eta} \trp
 e^{i\int_0^1 \dt (\hat B (t) +(\eta +i\dl)\hat N)}
\la{22}
\eea

\noi The latter equality is valid due to (\ref{19}).

The trace of ordered exponent in (\ref{22}) can be represented as
functional integral. In our case its explicit form  essentially
depends on some subtleties in its definition. So let me remind in a
few words the general construction of functional integral. For more
detailed discussion see, for instance, \cite{22}.

Let

\be
\check Z=\trp e^{\int^1_0 \dt \hat H}
\la{23}
\ee

\noi where $\hat H=\hat H(a^{\dag},a;t)$ is an operator in Fock
space ${\cal F}$.

In our case

\be
\hat H=(i\eta-\delta)\hat N +i\hat B
\la{24.1}
\ee

Equivalently, we can write

\be
\hat H=\hat H(\hat p,\hat q;t)
\la{24.2}
\ee

\noi

\be
\hat p=\frac{a+a^{\dag}}{\sqrt 2}, \qquad \hat q=
\frac{a-a^{\dag}}{i\sqrt 2}
\la{24}
\ee

Let $H=H(p,q;t)\equiv H(\bar z,z;t)$ be Weyl, or normal, or any
other symbol of the operator $\hat H$,

\be
z=\frac{p-iq}{\sqrt 2},\qquad \bar z=\frac{p+iq}{\sqrt 2}
\la{25}
\ee

Further, let $*$ be the operation that represent the multiplication
of operators in the language of symbols. This means that if $\hat
K=\hat K(\hat p,\hat q)$ and  $\hat L=\hat L(\hat p,\hat q)$ are
some operators with symbols $ K=K( p, q)$ and $ L= L( p, q)$ and
$\hat M=\hat K \hat L$ then

\be
M(p,q)=(K*L)(p,q)
\la{26}
\ee

In this terms

\be
\check Z=\lim_{n\to \infty}\int\mathop{dpdq}\left(
e^{\frac{1}{n}H(\cdot,\cdot;0)}*
e^{\frac{1}{n}H(\cdot,\cdot;\frac{1}{n})}*\ldots*
e^{\frac{1}{n}H(\cdot,\cdot;\frac{n-1}{n})}\right)(p,q)
 \la{27}
 \ee

The last formula can be rewritten as functional integral. Its
concrete form depends on the choice of a kind of symbols used
in (\ref{27}). In particular, for Weyl symbols

\be
\check Z={\cal N}\int_{PBC}DpDq\: e^{\int_0^1 dt\:(ip(t)\dot q(t)+
H_W(p(t),q(t);t))}
\la{28}
\ee

\noi whereas for normal symbols

\be
\check Z=\lime{\cal N}' \int_{PBC} DzD\bar z \: e^{
\int_0^1 \dt(\bar z(t)\dot z(t)+H_{norm}(\bar z(t),z(t+\e))}
\la{29}
\ee

\noi Here $PBC$ means 'periodic boundary conditions', $z$ and $\bar
z$ are {\rm independent} complex variables, and ${\cal N,N'}$ are
normalization constants (that we will usually omit in what
follows).

In our case

\bea
H_W(p,q;t)&=&H_W(z^{\dag},z;t)\nn\\
&=&(i\eta-\dl)z^{\dag}z+i\zd Bz-\frac{N}{2}(i\eta-\dl)
\la{30}
\eea

\noi where $p,q$ are connected with $z$ by formulae (\ref{25}),
and

\be
H_{norm}(\hat z,z)=(i\eta-\dl)\zd z +i\zd B z
\la{31}
\ee

Formulae (\ref{28}), (\ref{29}) correspond to standard sign
 conventions. However, for our purposes it is convenient to change
 variables

$$ z(t)\to z(1-t)$$

\noi Then formulae (\ref{28}), (\ref{29}) with symbols (\ref{30}),
(\ref{31}) can be rewritten as

\bea
\check Z&=&\int_{PBC} D^2 z\:\exp\left\{ i\int_0^1\dt \left[ i\zd (t)
\dot z(t) +\zd (t) B(t) z(t)\right.\right.\nn\\
& +&\left.\left.        (\eta
+i\dl)\zd(t)z(t)-\frac{N}{2}(\eta+i\dl)\right]\right\}
\la{32}
\eea

\noi where
$D^2z\equiv D({\rm Re} z) D({\rm Im} z)$ and

\bea
\check Z& =&\lime \int DzD\bar z \: \exp\left\{ i\int^1_0 \dt \left[
i\bar z(t) \dot z(t) +\bar z(t)B(t)z(t-\e)\right.\right.\nn\\
&{}&\left.\left. +(\eta+i\dl)\bar z(t)z(t-\e) \right] \right\}
\la{33}
\eea

\noi respectively.

There are several essential differences between formulae (\ref{32})
and (\ref{33}). First, in (\ref{32}) $z$ and $\zd$ are complex
conjugated variables whereas in (\ref{33}) $\bar z$ and $z$ are
independent. The second, the last term in the "action" in (\ref{32})
is absent in (\ref{33}). The third, there is the shift of "time"
variable in the last two terms in the "action" in (\ref{33}) that is
absent in (\ref{32}). In the next section we will show by explicit
calculations that this shift just compensates the absence of the
term

$$ -\frac{N}{2}(\eta +i\dl)$$

\noi    in (\ref{33}). Finally, it is worth to note that the limit
in (\ref{33}) must be evaluated {\it after} functional integration in
(\ref{33}) because this two operations don't commute. It will be
confirmed by explicit calculations in the section 3.

Substituting (\ref{32}) and (\ref{33}) in (\ref{22}) and evaluating
limits $\ep \to 0, \ \ \dl \to 0$ (bat not the limit $\e \to +0$!),
ones obtains, respectively,

\bea
Z&=&\int_{PBC} D^2z \: \de e^{i\int^1_0 \dt (i\zd\dot z+\zd Bz+
\eta(\zd z-1-\frac{N}{2})}
\la{34}
\eea

 \noi and

\bea
Z&=&\lime \int_{PBC} DzD\bar z \de \exp
\left\{ i\int_0^1\dt i\bar z
\dot z +\bar z B e^{-\e \frac{d}{dt}}z \right.
\nn\\
&+&\left.\eta(\bar ze^{-\e \frac{d}{dt}}z -1)\right\}
\la{35}
\eea

We won't try to justify here the validity of limiting procedure $\ep
\to 0$, $\mbox{$\dl \to 0$}$ because in the next section we will check
formulae (\ref{34}), (\ref{35}) by direct calculation.

In (\ref{34}) one can integrate with respect to $\eta$:

\be
Z=\Dzi
e^{i\int_0^1 \dt(iz^{\dag}\dot z +z^{\dag} Bz)}
\la{36}
\ee

But we cannot obtain $\dl$-function by integration with respect to
$\eta$ in (\ref{35}) because $\bar z$ and $z$ in (\ref{35}) are
independent complex variables.

One can also get another useful form of the representation
(       \ref{36}), namely

\be
Z=\int_{PBC}D^2z \: \prod \limits _t \dl(\zd (t)z(t)-1-\frac{N}{2})
e^{i\int_0^1 \dt(iz^{\dag}\dot z +z^{\dag} Bz)}
\la{37}
\ee

In what follows, we will sometimes omit symbol $\prod$ in formulae
of the type (\ref{37}).

To get formula (\ref{37}), it is sufficient to insert projectors
$\Pi_1$ represented in the form (\ref{21}) between each pair of
adjacent factors in (\ref{27}) and to repeat all calculations that
have led us to representation (\ref{36}).

The same arguments allow also to obtain an analogous variant of
(\ref{35}):

\bea
Z&=&\lime \int_{PBC} Dz(t)D\bar z(t)D\eta(t) \exp
\left\{ \right. i\int_0^1\dt i\bar z (t)
\dot z(t) +\bar z(t) B(t) e^{-\e \frac{d}{dt}}z(t)
 \nn\\
&+&\eta(t)(\bar z(t)e^{-\e \frac{d}{dt}}z(t) -1)\left.\right\}
\la{37.1}
\eea

Thus we have derived four different representations
(\ref{35})-(\ref{37.1}) for one quantity $Z$. They all appear to be
useful in quantum field theory.

 We have got (\ref{35})-(\ref{37.1}) assuming that $B\in su(N)$.
But these representation remains valid for every trace free matrix
by virtue of analytical continuation. Representation for $B\in
GL(N)$ can be obtained by extracting of the trace part of the matrix
$B$ at the beginning of calculations. Indeed, if

$$B=B'+ \frac{1}{N}I_N\tr B$$

\noi then $\tr B'=0$ and

\be
\trp e^{i\int_0^1 \dt B}=
e^{\frac{i}{N}\int_0^1 \dt \tr B}
\trp e^{i\int_0^1 \dt B'}
\la{39}
\ee

Another representation for matrices with non-zero trace will be
given in the next section (see eq. (\ref{58})).

Finally, let us derive a kind of representation (\ref{37}) for the
case

\be
B(t)=\sum_{j=1}^{M} C_j(t) \otimes D_j(t)
\la{39.1}
\ee

\noi where $C_j$ and $D_j$ are $N_1\times N_1$ and $N_2\times N_2$
matrices. Let $a_{(i)}^{r_i},\ \ a^{\dag}_{(i)r_i},\ \ i=1,2
\ \ r_i=1,2,\ldots ,N_i$ be two sets of annihilation and creation
operators that act in some Fock space, and $\Pi_{1\otimes 1}$ is a
projector on the space

\be
{\cal H}_{1\otimes 1}=\left\{\phi^{r_1 r_2}a^{\dag}_{(1)r_1}
a^{\dag}_{(2)r_2}|0>\right\}
\la{39.2}
\ee

One can write

\bea
\lefteqn{
\trp e^{i\int_0^1 \dt B}}\nn\\
&&=\tr \Pi_{1\otimes 1}\: \mbox{P}\: \exp\left\{ i\int^1_0 \dt
\sum\limits_{j=1}^M(a^{\dag}_{(1)}C_j a_{(1)})
(a^{\dag}_{(2)}D_j a_{(2)})\right\}
\la{39.3}
\eea

\noi The last formula ia an analog of (\ref{18}). Then, repeating the
calculations that have been done for derivation of eq. (\ref{36})
from eq. (\ref{18}), one obtains:

\bea
\trp e^{i\int_0^1 \dt B}
&=&\int D^2z_1 D^2 z_2 \prod\limits_{i=1}^2
\dl (z^{\dag}_{(i)}z_{(i)}-1-\frac{N}{2})
\nn\\
&&\exp \left\{ i\int_0^1 \dt \left[ i\sum\limits_{i=1}^2
z^{\dag}_{(i)}\dot z_{(i)}\right.\right.\nn\\
&&\left.\left.  +\sum\limits_{j=1}^M
(z^{\dag}_{(1)}C_j z_{(1)} -\frac{1}{2}\tr C_j)
(z^{\dag}_{(2)}D_j z_{(2)}-\frac{1}{2}\tr D_j)
\right] \right\}
\la{39.4}
\eea

Similar arguments allow to obtain the analog of eqs. (\ref{35}),
(\ref{37}), and (\ref{37.1}) for an matrix $B$ defined by eq.
(\ref{39.1}).

\section{Alternative proof of bosonic path integral representation
  for the trace of path ordered exponent}

At first, we will evaluate the integral (\ref{36}) assuming that
$B\in su(N)$  . One can write:

\bea
Z&=&\lime \Dzi e^{-\e\int^1_0 \dt \zd z}e^{i\intt \dt(i\zd \dot z
+\zd B z)}\nn\\
&=&\lime \de e^{-i\eta\left( 1+\frac{N}{2}\right)}\int_{PBC} D^2z \:
e^{\intt \dt \zd\left(-\frac{d}{dt} +iB+i\eta -\e\right)z}
\la{40}
\eea

Performing functional integration, one gets:

\be
Z=\lime \de \frac{e^{-i\eta\left( 1+\frac{N}{2}\right)}}{
{\det}\left(-\frac{d}{dt} +i(B+\eta)-\e\right)_{PBC}}
\la{41}
  \ee

To evaluate integral in (\ref{41}), let us consider eigenvalue
problem

\be
\left(-\frac{d}{dt} +i(B+\eta)-\e\right)\phi (t)=\lm \phi(t)
\la{42}
\ee

\be
\phi(0)=\phi(1)
\la{43}
\ee

The general solution of eq. (\ref{42}) is

\be
\phi=e^{(-\lm +i\eta -\e)t} U(t)\chi
\la{44}
\ee

\noi where $U(t)$ is a solution of (\ref{15}) and vector $\chi$
doesn't depend on $t$.

Let $e^{i\al_r}$ and $\xi_r,\ \ r=1,\ldots,N$ be eigenvalues and
eigenvectors of the matrix $U(1)$:

\be
U(1)\xi_r=e^{i\al_r}\xi_r
\la{45}
\ee

One notes that $U(1)\in SU(N)$ and so $\al_r$ are real and

\be
\sum_{r=1}^N \al_r=0
\la{46}
\ee

Further, to satisfy (\ref{43}) we must put

\be
\lm\equiv \lm_{rn}=-\e +i\eta +i\al_r +2\pi i n,\ \ n=0,\pm 1,\ldots
\la{47}
\ee

\noi and $\chi=\xi_r$. So

\bea
\lefteqn{{\det}\left(-\frac{d}{dt} +i(B+\eta)-\e\right)_{PBC}=
\prod\limits_{r=1}^{N}
\prod\limits_{n=-\infty}^{\infty}(2\pi i n+i\al_r +i\eta-\e)}
\nn\\
&=&\left[-i\prod_{n=-\infty}^{\infty}(2\pi in)\right]^r
\prod\limits_{r=1}^{N}
(\al_r+\eta +i\e)\prod\limits_{n\ne 0}
\left(1+\frac{\al_r +\eta +i\e}{2\pi n}\right)\nn\\
&=&\left[-i\prod_{n=-\infty}^{\infty}(2\pi in)\right]^r
\prod\limits_{r=1}^{N}(\al_r+\eta +i\e)
\prod\limits_{n=1}^{\infty}
\left( 1-\frac{(\al_r+\eta +i\e)^2}{4\pi^2n^2}\right)
\la{48}
\eea

\noi Then, omitting irrelevant infinite constant and using well-known
formula

\be
\prod\limits_{n=1}^{\infty}\left(
1-\frac{a^2}{n^2}\right)=\frac{1}{\pi a} \sin \pi a \la{49} \ee

\noi one gets

\be
{\det}\left(-\frac{d}{dt}
+i(B+\eta)-\e\right)_{PBC}=
\prod_{r=1}^{N}\sin \left(
\frac{\al_r+\eta+i\e}{2}\right)
\la{50}
\ee

Substituting (\ref{50}) in (\ref{41}), one obtains contour integral

\be
Z=\lime \de \frac{e^{-i\eta\left(1+\frac{N}{2}\right)}}{
\prod_{r=1}^{N}\sin \left(
\frac{\al_r+\eta+i\e}{2}\right)
}
\la{51}
\ee

One can close the contour of integration in (\ref{51}) in the lower
half plane and represent $Z$ as the sum of residues in the poles

\be
\eta_{rn}=-\al_r -i\e +2\pi n
\la{52}
\ee

The contribution of the pole with given $r$ and $n$ is

\be
2\pi i(-1)^{n+1}\frac{e^{i(\al_r+i\e)}}{
\prod\limits_{{1\le s\le N \atop s\ne r}}\sin \left(
\frac{\al_s-\al_r}{2}\right)
}
\la{53}
\ee
\noi So, omitting again inessential constant, one gets:

\bea
Z&=&\lime \sum\limits_{r,n}
\left(
\begin{array}{c}
\mbox{contribution of the residue in}\\
\eta_{rn}=-\al_r -i\e +2\pi n
\end{array}
\right)
\nn\\
&=&
\sum\limits_{r=1}^N\frac{e^{i\al_r(1+\frac{N}{2})}}{
\prod\limits_{
{\scriptstyle {s=1 \atop  s\ne r}}
}^N
\sin \left( \frac{\al_s-\al_r}{2} \right)}
\la{54}
\eea

\noi Finally, using elementary but rather non-obvious identity

\be
\sum\limits_{r=1}^N\frac{e^{i\al_r(1+\frac{N}{2})}}{
\prod\limits_{
{\scriptstyle
{s=1\atop s\ne r}}}
^N
\sin \left( \frac{\al_s-\al_r}{2} \right)}
=
(-2i)^{N-1}
\left(e^{\frac i 2 \sum\limits_{r=1}^N \al_r}\right)
\sum\limits_{r=1}^N e^{i\al_r}
\la{55}
\ee

\noi (that is valid for any complex numbers $\al_r$; see Appendix A
for proof), one obtains:

\be
Z=
\sum\limits_{r=1}^N e^{i\al_r}=\tr U(1) =\trp e^{i \intt \dt B(t)}
\la{56}
\ee
\noi This proves the representation (\ref{36}).

Throughout the proof we didn't control inessential numerical
normalization factors arising in front of functional integrals,
determinants, etc. They can be easily reconstructed from
normalization condition

\be
Z\raisebox{-10pt}{\rule{0.4pt}{20pt}${\scriptstyle B=0}$}
=N
\la{57}
\ee

The representation (\ref{36}) is valid for traceless matrix $B$. In
general case the following representation is valid

\be
Z=e^{-\frac i 2 \intt \dt \tr B(t)}
\int D^2z \: \dl \left( \intt \dt \zd z -1- \frac N 2\right)
e^{i\int_0^1 \dt(iz^{\dag}\dot z +z^{\dag} Bz)}
\la{58}
\ee

\noi This representation is alternative to one given by eq.
(\ref{39}).

To prove (\ref{58}), it is sufficient  to repeat the proof given for
representation (\ref{36}) but without using the condition (\ref{46})
. One can trace the cancellation of the pre-integral factor
in (\ref{58}) and contribution of the factor

$$
e^{\frac i 2 \sum\limits_{r=1}^N \al_r}
$$

\noi in (\ref{55}).

Now let us check the representation (\ref{37}).
Representing $\dl$-function in (\ref{37}) as

\be
\prod_t \dl \left( \zd z -1 -\frac N 2 \right)=\int
D\eta \: e^{i\intt \dt
\left( \zd z -1 -\frac N 2 \right)\eta
}
\la{60}
\ee

\noi and integrating over $\zd, z$, one obtains the analog of
(\ref{41}):

\be
Z=\lime \int D\eta(t)\: \frac{e^{-i\left( 1+\frac{N}{2}\right)
\intt \dt \eta(t)}}{
{\det}\left(-\frac{d}{dt} +i(B(t)+\eta(t))-\e\right)_{PBC}}
\la{61}
  \ee

The eigenvalue problem

\be
\left(-\frac{d}{dt} +i(B(t)+\eta(t))-\e\right)\phi (t)=\lm \phi(t)
\la{62}
\ee

\be
\phi(0)=\phi(1)
\la{62.1}
\ee

\noi has the solution

\be
\lm\equiv \lm_{rn}=-\e +i\al_r
+i\intt \dt \eta(t)
 +2\pi i n,\ \ n=0,\pm 1,\ldots
\la{63}
\ee

\noi where numbers $\al_r$ are defined by equation (\ref{45}). So
determinant in (\ref{61}) depends on $\eta(t)$ only via

\be
\eta=\intt \dt \eta(t)
\la{64}
\ee

\noi Therefore, if one introduce new variables $\eta$ and

\be
\eta_n=\intt \dt e^{2\pi int} \eta(t),\ \ n\ne 0
\la{65}
\ee

\noi instead of $\eta(t)$ in (\ref{61}), one finds that integration
with respect to $\eta_n$ gives only inessential constant and so the
integral (\ref{61}) transforms just in the integral (\ref{41}). But
this means that we reduced the proof of representation (\ref{37}) to
one of the representation (\ref{36}) which we have already proved.

Finally, let us prove the validity of the representations (\ref{35})
and (\ref{37.1}). The proof can be reduced again to one of the
representation (\ref{36}) by means of the formula

\be
\lime {\det} \oee =e^{\frac i 2 \intt \dt \tr B(t)}
{\det} \left( -\frac{d}{dt} +iB(t) \right)
\la{66}
\ee

\noi that is valid for any $N\times N$ matrix $B(t)$ (the proof of
(\ref{66}) is given in Appendix~B). Indeed, evaluating the integral
with respect to $\bar z, \, z$ in (\ref{35}), one obtains, after
inserting of the factor $\exp\left( -\dl \intt \dt \zd
(t)z(t+\e)\right)$,

\be
Z=\lim_{\dl \to +0}\lime \de \frac{e^{-i\eta
}}{ {\det}\left(-\frac{d}{dt}
+i(B+\eta+i\dl)e^{-\e\frac{d}{dt}}\right)_{PBC}}
\la{67}
\ee

\noi and further application of (\ref{66}) reduce (\ref{67}) to
(\ref{41}). (Remind, that $\tr B=0$ by assumption.)

The same arguments allow to check the representation(\ref{37.1}).

Thus we have checked all four representations for the trace of path
ordered exponent derived in the section 2. We see the proofs given
in the present section are very unlike ones given in the  section 2.
This can be considered as strong confirmation of the validity of
the representations
obtained.

\section{A variant of non-Abelian Stokes theorem}

Analogs of Stokes theorem for Wilson loop

\be
Z=\wl
\la{68}
\ee

\noi were proposed by several authors \cite{23}. In this works
Wilson loop is expressed via some area integral over surface spanned
on $\g$ with rather complicated path ordered prescriptions. Recently
Dyakonov and Petrov derived another, very smart variant of
non-Abelian Stokes theorem. Their result we already
cited (see (\ref{14})).

In this section we will prove a new variant of non-Abelian Stokes
theorem for $A_{\mu}\in su(N)$ that is similar to one given by
Dyakonov and Petrov. For the case $N=2$ we will be able to derive
from our results the Dyakonov-Petrov's formula (\ref{14}) but in
slightly corrected form.  The little discrepancy between our results
and those by Dyakonov and Petrov arises, most likely, because of some
subtleties in the definition of the corresponding functional
integrals.

Let the path $\gamma$ in (\ref{68}) be parametrized as

\be
\g=\{ x^{\mu}=q^{\mu}(t),\:0\le t\le 1,\: q^{\mu}(0)=q^{\mu}(t)\}
\la{69}
\ee

\noi Then

\be
Z=\trp e^{i\intt \dt \dot q ^{\mu}(t) A_{\mu}(q(t))}
\la{70}
\ee

\noi and we can apply the representation (\ref{37}):

\be
Z=
\int D^2\psi \dl (\zd z -1- \frac N 2)
 e^{\intt \dt \zd\left(-\frac{d}{dt} +i\dot q A\right)z}
\la{71}
\ee

Let $\xi^r=\xi^r(x), \: r=1,\ldots,N$ be any field such that

\be
\xi^r(q(t))=z^r(t)
\la{72}
\ee

\noi The formula (\ref{71}) can be rewritten as

\be
Z=
 \int \mathop{D^2\xi} \prod\limits_{x\in\g}\dl\left( \xi^{\dag}(x)
\xi(x)-1-\frac N 2 \right)
e^{-\oint_{\g} dx^{\mu} \: \xi^{\dag}D_{\mu}\xi(x)}
\la{73}
\ee

\noi where $D_{\mu}=\partial_{\mu}-iA_{\mu}$ is the usual covariant
derivative. Further,
applying the classical Stokes theorem, one can easy to prove that

\be
\oint_{\g} dx^{\mu} \: \xi^{\dag}D_{\mu}\xi(x)
=
\int_{\Si} \ddx \left(D_{\mu}\xi^{\dag}D_{\nu}\xi -\frac i  2
\xi^{\dag}F_{\mu\nu}\xi \right)
\la{74}
\ee

\noi where $\Si$ is any surface for which $\partial \Si=\g$.

Substituting (\ref{74}) in (\ref{73}), we get our variant of
non-Abelian Stokes theorem:

{\samepage
\bea
Z
\equiv&&\wl\nn\\
=&&
 \int \mathop{D^2\xi} \prod\limits_{x\in\g}\dl\left( \xi^{\dag}(x)
\xi(x)-1-\frac N 2 \right)
e^{
\int_{\Si} \ddx \left(-D_{\mu}\xi^{\dag}D_{\nu}\xi +\frac i  2
\xi^{\dag}F_{\mu\nu}\xi \right)
}\nn\\
\la{75}
\eea

}

Now let us transform eq. (\ref{75}) into the form that would be
similar to eq. (\ref{14}).

Let $\lm^a,\: a=1,\ldots,N^2-1$ be Hermitian generators of $SU(N)$
in the fundamental representation. They can be normed as

\be
\tr \lm^a\lm^b=2\dl^{ab}
\la{76}
\ee

\noi and satisfy equations

\be
[\lm^a, \lm^b] =2 i f^{abc}\lm^c
\la{77}
\ee
\be
[\lm^a, \lm^b]_{+} = \frac 4 N \dl^{ab}I_N +2d^{abc}\lm^c
\la{78}
\ee

One notes that

\be
A_{\mu}=\frac{\lm^a}{2}A_{\mu}^a,\: \ F_{\mu\nu}=
\frac{\lm^a}{2}F_{\mu\nu}^a=
\frac{\lm^a}{2}(\partial_{\mu}A^a_{\nu}-
\partial_{\nu}A^a_{\mu} +f^{abc}A^b_{\mu}A^c_{\nu})
\la{78'}
\ee

Matrices $\lm^a$ also obey "Fierz" identities:

\bea
\frac 1 2 \lm^a_{ij}\lm^a_{kl}+
\frac 1 N \dl_{ij}\dl_{kl}&=&
\dl_{il}\dl_{kj}
\la{79}\\
if^{abc}\lm^a_{ij}\lm^b_{kl}\lm^c_{mn}&=&
2(\dl_{il}\dl_{mj}\dl_{km}-
\dl_{in}\dl_{jk}\dl_{ml})
\la{80}
\eea

One introduces new variables

\be
I^a(x)=\xi^{\dag}(x)\lm^a\xi(x)
\la{81}
\ee

Variables $I^a$ are not independent. Using "Fierz" identity
(\ref{79}), formula (\ref{81}) can be represented as

\bea
\frac{1}{\xi^{\dag} \xi}(I^a\lm^a)_{ij}+\frac 1 N \dl_{ij}=
\xi_i\xd_j\frac{1}{(\xd\xi)}
\la{82}
\eea

Let $I\equiv I^a\lm^a, \: \xd\xi\equiv c$. One notes that
$c=1+\frac N 2$ on the surface of integration in (\ref{75}).

The matrix $I$ can be represented in the form

\be
I=U diag(a_1,\ldots,a_N)U^{\dag}
\la{83}
\ee

\noi where

\be
\sum_{i=1}^{N}a_i=0,\ \ U\in SU(N)
\la{84}
\ee

The matrix in R.H.S. of eq. (\ref{82}) is a projector on the one
dimensional subspace.  This means that eigenvalues of the matrix in
the L.H.S.  of eq.  (\ref{82}) are all equal to zero except only one
that is equal to 1.  So, up to renumbering of eigenvalues,

\bea
a_i&=&-\frac{2c}{N},\ \ i=1,\ldots,N-1\nn\\
a_N&=&2c\frac{(N-1)}{N}
\la{85}
\eea

One notes that

\bea
\tr I^k&=&(N-1)\left( -\frac{2c}{N}\right)^k +\left[
2c\frac{(N-1)}{N}\right]^k\nn\\
&=&\frac{(2c)^k (N-1)}{N^k}
\left[ (N-1)^{k-1} +(-1)^k
\right]
\la{94}
\eea

Thus the matrix $I$ has $N-1$ coincident eigenvalues. But this means
that the solutions of eq. (\ref{82}) are in one-to-one
correspondence with the points of a \co $\approx CP^{N-1}$.

Indeed, the matrix $U$ in formula (\ref{83}) can be presented in the
form

\be
U=U_1V_1V_2
\la{85.1}
\ee

\noi where

\bea
V_1=
\left(
\begin{array}{cc}
\tilde{V}_1&
\begin{array}{c}
0\\
\vdots\\ 0
\end{array}\\
0\ldots 0&1
\end{array} \right), \ \ \tilde V_1 \in SU(N-1)
\la{85.2} \\
V_2=\exp \{ diag(t,\ldots,t,-(N-1)t \}\nn
\eea

\noi and a matrix $U_1$ represents an element of the \co . So, due to
coincidence of the first $N-1$ eigenvalues of the matrix $I$, one
finds

\be
I=U_1 diag\left(-\frac{2c}{N},\ldots,-\frac{2c}{N},
2c\frac{(N-1)}{N}\right) U^{\dag}_1 \la{86} \ee

\noi Therefore the set of solutions of eq. (\ref{82}) is isomorfic to
\co .

Further, using identities (\ref{79}), (\ref{80}), one finds that on
the surface $\xd \xi$$=c=const$

\bea
D_{\mu}\xd D_{\nu}\xi&=&\frac{1}{8c^2}\tr ID_{[\mu}ID_{\nu]}I
\la{87}\\
\xd F_{\mu\nu}\xi&=&\frac 1 2 \tr IF_{\mu\nu}
\la{88}
\eea

\noi where

\be
D_{\mu}I=\partial_{\mu}I-i[A_{\mu},I]
\la{89}
\ee

Now let us insert in (\ref{75}) an identity

\be
1=\prod_{x\in\Si}\int\prod_a dI^a(x)\dl(\xd(x) \lm^a \xi(x) -I^a(x))
\la{90}
\ee

\noi By virtue of (\ref{87}), (\ref{88}), the eq. (\ref{75}) can be
rewritten in the form

\be
Z=\int D\mu(I)\exp\left\{
-i\int\limits_{\Si}\ddx\left[
\frac{1}{8c^2}\tr ID_{[\mu}ID_{\nu]}I-\frac 1 4
\tr IF_{\mu\nu}\right]\right\}
\la{91}
\ee

\noi where

\be
D\mu(I)=\left(\prod_a DI^a\right)
\int D^2\xi \dl(\xd \xi- 1-\frac N 2)\prod_a\dl(\xd \lm^a \xi-I^a)
\la{92}
\ee

The consideration given above shows that
$D\mu(I)$ is nothing but $SU(N)$-invariant measure on the
\co{}

The explicit form of the measure $D\mu(I)$ is rather cumbersome but,
fortunately, it appears that measure $D\mu(I)$ in (\ref{91}) can be
replaced by the measure

\be
D\mu'(I)\equiv \left(DI^a\right) \prod_{k=2}^N \dl(\tr I^k- c_{k,N})
\la{93}
\ee

\noi where numbers $c_{k,N}$ are defined by formula (\ref{94}) in
which one must put $2c=N+2$ by virtue of the first $\dl$-function in
eq.(\ref{92}).

 Indeed, numbers $\tr I^k$ define uniquely
characteristic equation for the matrix $I$ and, consequently, its
eigenvalues. This leads to representation (\ref{86}). So on the
surface $\tr I^k=c_{k,N}$ any functional $\Phi(I)$ is invariant under
transformation

\be
\Phi(I)\to \Phi(V_1 V_2 IV^{\dag}_2V^{\dag}_1)
\ee

\noi where a matrix $V_1 V_2$ is defined by eqs. (\ref{85.2}).
Therefore in any integral

\be
\int D\mu'(I)\Phi(I)
\ee

\noi one can perform all integrations except those corresponding to
integration over \co :

\be
\int D\mu'(I)\Phi(I)
=const \int D\mu(I)\Phi (U_1diag(-\frac{2c}{N},\ldots,-\frac{2c}{N},
2c\frac{N-1}{N})U^{\dag}_1)
\ee

\noi (see (\ref{86})). But this just means, in particular, that one
can replace the measure $D\mu(I)$ in (\ref{91}) by the measure
$D\mu'(I)$ defined by eq. (\ref{93}).

Now we can formulate our final result:

\bea
\lefteqn{
\trp \exp\left\{i\oint_{\gamma} \dxm A_{\mu} \right\}}\nn\\
&=&\int DI\prod_{k=2}^N \dl(\tr I^k-c_{k,N})\nn\\
&&\exp \left\{
i\int_{\Si}\ddx\left[
-\frac{1}{2(N+2)^2}\tr (ID_{\mu}ID_{\nu}I) +\frac 1 4 \tr IF_{\mu\nu}
\right]\right\}\nn\\
&&\la{94'}
\eea

\noi where

\be
c_{k,N}=\frac{(N-1)(N+2)^k}{N^k}\left[ (N-1)^{k-1} +(-1)^k
\right]
\la{95}
\ee

Let us compare our results with those due to Dyakonov and Petrov
\cite{21}. Putting in (\ref{94'})  $N=2,\ \ I=I^a\sigma^a$ and the
changing $I^a\to 2I^a$ , we get:

\bea
\lefteqn{
\trp \exp\left\{i\oint_{\gamma} \dxm A_{\mu} \right\}}\nn\\
&=&
\int DI \dl(I^2-1)\exp\left\{-\frac i 2 \int \ddx
\left( \ep^{abc}I^aD_{\mu}I^b D_{\nu}I^c-I^a F^a_{\mu\nu}
\right)\right\}\nn\\
&&\la{97}
\eea

Comparing formulae (\ref{97}) and (\ref{14}), we see that they
differ by the factor 1/2 in front of the "action". Most likely this
discrepancy arises because of some subtleties in the definition of
the functional integrals that play the important role in our
discussion. In particular, if one ignores the difference
between functional integrals constructed by means of Weyl and normal
symbols, one obtains the following representation instead of
(\ref{73}):

\be
Z=
 \int \mathop{D^2\xi} \prod\limits_{x\in\g}\dl\left( \xi^{\dag}(x)
\xi(x)-1 \right)
e^{-\oint_{\g} dx^{\mu} \: \xi^{\dag}D_{\mu}\xi(x)}
\la{98}
\ee

\noi Further, using formulae (\ref{87}) and (\ref{94}) with $c=1$
instead of $c=1+\frac N 2$, one can easy trace that representation
(\ref{98}) leads exactly to Dyakonov-Petrov formula (\ref{14}). But
representation (\ref{98}) is wrong. So just formula (\ref{97}) must
be considered as correct version of non-Abelian Stokes theorem for
$N=2$.

This doesn't mean, however, that Dyakonov-Petrov formula (\ref{14})
is incorrect. But it means that the definition of functional
integral in (\ref{14}) must be clarified.

Another discrepancy between our results and those due to Dyakonov
and Petrov arises in the case $N\ge 3$. Dyakonov and Petrov pointed
out in their work \cite {21} that integration in the functional
integral representation for Wilson loop must be performed over the
coset $SU(N)/[U(N)]^{N-1}$
whereas in our representation (\ref{94'}) integration is carried out,
in fact, over the \co . But nowadays it is hard to discuss this
discrepancy because no explicit formulae for the case $N\ge 3$ were
given in \cite{21}.

{\hsize=13.26cm
{\section{Bosonic worldline path integral representation for
fermionic  determinants and
Green functions in Euclidean space}}

Bosonic worldline path integral representation for fermionic
determinants can be obtained directly from formulae (\ref{8}),
(\ref{10}) and results of the section 2:

\bea
\lefteqn{
\ln{\det}(i\nh+im)}\nn\\
&=&\dT\dQ\nn\\
&&
{\cal N}_T\exp\left\{
\intt \dt \left[ -\frac{\dot q^2}{4T}-\zd \dot z
-\psi^{\dag}\dot{\psi}+i\dot q \zd A_{\mu} z\right.\right.\nn\\
&&\left.\left.
-T(\psi^{\dag}\sigma^{\mu\nu}\psi)(\zd F_{\mu\nu} z)
\vphantom{
\intt \dt \left[ -\frac{\dot q^2}{4T}-\zd \dot z\right.
}
\right] \right\}
\la{99}
\eea

Here variables $z=\{z^r, \ \ r=1,\ldots,N\}$ and $\psi=\{\psi^i,\:
i=1,,2,3,4\}$ describe colour and spin degrees of freedom
respectively, $\zd$ and $\psi^{\dag}$ are complex conjugated to $z$
and $\psi$,

\bea
D^2 z&\equiv&\prod\limits_t \prod\limits_{r=1}^N
\mathop{d(\mbox{Re}\, z^r(t))}\mathop{d(\mbox{Im}\, z^r(t))}\la{100}\\
D^2 \psi&\equiv&\prod\limits_t \prod\limits_{r=1}^N
\mathop{d(\mbox{Re}\, \psi^i(t))}\mathop{d(\mbox{Im}\, \psi^i(t))}
\la{100.1}
\eea

\noi and ${\cal N}_T$ is a normalization constant. The latter can be
evaluated from the condition

\be
<x|\tr
e^{-T\wH}|x>\raisebox{-10pt}{\rule{0.4pt}{20pt}${\scriptstyle A=0}$}=
4N<x|e^{-T\partial_{\mu}\partial_{\mu}}|x>=\frac{N}{2\pi^2T^2}
\la{101}
\ee

Indeed, putting $A=0$ in (\ref{99}) and comparing the result with
(\ref{101}), one obtains:

\bea
{\cal N}_T^{-1}
&=&\left( \frac{N}{2\pi^2T^2}\right)^{-1}
\int\limits_{q(0)=q(1)=x} Dq\:\int_{PBC}D^2\psi D^2 z\delta\left(
z^{\dag}z -1-\frac N 2 \right)\nn\\
&&\delta\left( \psi^{\dag}\psi
-3\right)
\exp\left\{
\intt \dt \left[ -\frac{\dot q^2}{4T}-\zd \dot z
-\psi^{\dag}\dot{\psi}\right]\right\}
\la{102}
\eea

\noi Obviously, ${\cal N}_T$ doesn't depend on $x$.

Remind, that in Euclidean space $\psi^{\dag}\psi$ is $SO(4)$ scalar.
So representation (\ref{99}) is manifestly relativistic and gauge
invariant.

Our next task is to derive bosonic worldline path integral
representation for Euclidean Green functions. In what follows, we
restrict ourselves only to derivation of such representation for
generating functional $Z(j)$ for vacuum correlators

\be
<\psi^{\dag}_{f_1}(x_1)\psi_{f_1}(x_1)\ldots
\psi^{\dag}_{f_n}(x_n)\psi_{f_n}(x_n)>
\la{103}
\ee

\noi However, our method is quite general and can be easily applied
to derivation of analogues representations for arbitrary Green
functions.

Standard functional integral representation for $Z(j)$ can be
written as

\be
Z(j)=\int DA\: e^{S_{YM}}\int D\Psi D\bar{\Psi}
e^{\left\{ \sum_f\int dx \:\left[ i\bar{\Psi}_f \nh\Psi_f+im_f
\bar{\Psi}_f\Psi_f-ij_f\bar{\Psi}_f\Psi_f\right]\right\}}
\la{104}
\ee

Integrating with respect to fermionic fields, we get

\be
Z(j)=\int DA\: e^{S_{YM}}\prod_f{\det}(i\nh+im_f-ij_f)
\la{105}
\ee

So our task is reduced to derivation of bosonic path integral
representation for determinant

$$
{\det}(i\nh+im-ij)
$$

The latter problem can be easily solved by applying of the
results obtained in the section 2. Indeed,

\bea
\lefteqn{
\ln{\det} (i\nh+im-ij)=\frac 1 2
 \ln[{\det}(i\nh+im-ij)\g^5]^{2}}\nn\\
&=&\frac 1 2
\ln{\det}(\nabla_{\mu}\nabla^{\mu}-\sigma^{\mu\nu}F_{\mu\nu}
+\hat{\partial}j+(m-j)^2)\nn\\
&=&\dT\tr e^{-T
(-\nabla_{\mu}\nabla^{\mu}+\sigma^{\mu\nu}F_{\mu\nu}-\hat{\partial}
j+2mj-j^2)}
\la{105'}
\eea

Using representation for path ordered exponent from section 2, one
obtains:

\bea
\lefteqn{\ln{\det}(i\nh+i(m+j))}\nn\\
&=& \dT\dQ \nn\\
&&\exp\left\{
\intt \dt \left[ -\frac{\dot q^2}{4T}-\zd \dot z
-\psi^{\dag}\dot{\psi}+i\dot q \zd A_{\mu} z\right.\right.\nn\\
&&\left.\left.-T(\psi^{\dag}\sigma^{\mu\nu}\psi)(\zd F_{\mu\nu} z
 -\psi^{\dag}\g^{\mu}\psi\partial_{\mu}j(q)-2mj(q)+j^2(q) \right]
 \right\}
 \la{106}
 \eea

 Substituting (\ref{105'}) in (\ref{106}) for each $f$, we get
 worldline bosonic path integral representation for generating
 functional $Z(j)$. The corresponding formulae for $n$-point
 correlators are rather cumbersome but quite computable. Author
 hopes that they can be used for computer simulations on the
 lattice.

{\hsize=13.26cm
\section{Bosonic worldline path integral
representation for fermionic determinants and Green
functions in Minkowski space and quasiclassical approximation in QCD}
}

The derivation of bosonic worldline path integral representation for
fermionic determinants in Minkowski space is slightly more involved
then one in Euclidean space. The origin of complications is
non-unitarity of finite dimensional representation of the Lorentz
group.

The analog of the representation (\ref{8}) in Minkowski space can be
written as

\be
\ln{\det}(i\nh-m)=\dTM \tr e^{iT\wH}
\la{107}
\ee

Trace in (\ref{107}) can be represented as functional integral:

\be
\tr e^{iT\wH}=\int_{PBC} Dq \: \trp
e^{i\intt \dt \left( -\frac{\dot q^2}{4T}+\dot q A(q)+T
\sigma^{\mu\nu}F_{\mu\nu}(q)\right)}
\la{108'}
\ee

\noi eq. (\ref{108'}) is an analog of eq. (\ref{10}). Path ordering in
(\ref{108'}) corresponds to colour and spinor structures.

However, in contrast to Euclidean case, we cannot directly use the
representations of the type (\ref{37}), (\ref{39.4}) to write
ordered exponent in (\ref{108'}) as functional integral. Indeed,
those representation comprise, in particular, the factor

\be
\dl(\pdd\psi-3)
\la{108''}
\ee

\noi (see (\ref{99})) that is not Lorentz invariant because spinor
representations of Lorentz group are not unitary.

To obtain manifestly Lorentz invariant representation, we will use,
at first, representation (\ref{37}) for describing of colour degrees
of freedom and representation (\ref{37.1}) for describing of spinor
ones. In such terms eq. (\ref{108'}) can be rewritten as

\bea
\lefteqn{\tr e^{i\wH}}\nn\\
&=&\lime \int_{PBC}DqD^2zD\bar{\psi}D\psi D\lm \dl(\zd z-1-\frac N 2)
\nn\\
&&{\cal N}_T \exp\left\{i\intt\dt\left[-\frac{\dot q^2}{4T}+i\zd z
+i\bar{\psi}\dot{\psi}+\right.\right.\nn\\
&&\left.\left.\zd \dot q A(q) z+ T(\bar{\psi}\sigma^{\mu\nu}e^{-\e
\frac d{dt}}\psi)(\zd F^{\mu\nu}z)+\lm(\bar{\psi}e^{-\e\frac{d}{dt}}
\psi-1)\right]\right\}
\la{108}
\eea

In eq. (\ref{108}) the measure $D^2z$ is defined by (\ref{100}) but
$\psi$ and
$\bar{\psi}$ are independent complex variables. ${\cal N}_T$ is a
normalization constant that will be computed later.

Integrating over $\bar{\psi},\ \ \psi$ in (\ref{108}), one gets:

\bea
\lefteqn{\tr e^{i\wH}}\nn\\
&=&\lime \int_{PBC}DqD^2z D\lm \dl(\zd z-1-\frac N 2)
\nn\\
&&{\cal N}_T
{\det}^{-1}\left[-\frac d{dt}+iT\sigma^{\mu\nu}(\zd F_{\mu\nu}z)
e^{-\e\frac d{dt}}+i\lm e^{-\e\frac d{dt}}\right]
\nn\\
&& \exp\left\{i\intt\dt\left[-\frac{\dot q^2}{4T}+i\zd \dot z+
\zd \dot q A(q) z- \lm \right]\right\}
\la{109}
\eea

But

\bea
\lefteqn{\lime
{\det}^{-1}\left[-\frac d{dt}+iT\sigma^{\mu\nu}(\zd F_{\mu\nu}z)
e^{-\e\frac d{dt}}+i\lm e^{-\e\frac d{dt}}\right]}\nn\\
&=&e^{-2i\intt \dt \lm (t)}
{\det}^{-1}\left[-\frac d{dt}+iT\sigma^{\mu\nu}(\zd F_{\mu\nu}z)
+i\lm\right]
\la{110}
\eea

\noi by virtue of identity (\ref{66}). Further,

\bea
\lefteqn{
{\det}^{-1}\left[-\frac d{dt}+iT\sigma^{\mu\nu}(\zd F_{\mu\nu}z)
+i\lm\right]}
\nn\\
&=&{\det}^{-1}\left[-\gamma^0\frac d{dt}+\gamma^0iT\sigma^{\mu\nu}(\zd
F_{\mu\nu}z) +i\g^0\lm\right]
\la{111}
\eea

\noi The operator in R.H.S. of (\ref{111}) is {\it anti-Hermitean}.
So we can write

\bea
\lefteqn{
{\det}^{-1}\left[-\gamma^0\frac d{dt}+\gamma^0iT\sigma^{\mu\nu}(\zd
F_{\mu\nu}z) +i\g^0\lm\right]}
\nn\\
&=&\int_{PBC}D^2\psi e^{i\intt \dt (i\pdd \g^0\dot{\psi}+T(\pdd\g^0
\sigma{\mu\nu}\psi)(\zd F_{\mu\nu}z)+\lm\pdd\g^0\psi)}
\la{112}
\eea

In the last formula $\pdd$ and $\psi$ are already complex conjugate
variables, measure $D^2\psi$ is defined by eq. (\ref{100.1}), and the
"action" is real. So functional integral (\ref{112}) is well defined.

Introducing standard notations $\bar{\psi}=\pdd\g^0$ and substituting
(\ref{110})-(\ref{112}) in (\ref{109}), one obtains, after
integration with respect to $\lm$,

\bea
\lefteqn{
\tr e^{iT\wH}}\nn\\
&=&\dQM \nn\\
&&{\cal N}_T \exp\left\{i\intt\dt\left[-\frac{\dot q^2}{4T}+i\zd z
+i\bar{\psi}\dot{\psi}+\right.\right.\nn\\
&&\left.\left.\zd \dot q A(q) z+ T(\bar{\psi}\sigma^{\mu\nu} \psi)(\zd
F^{\mu\nu}(q)z)\right]\right\}
\la{113}
\eea

The normalization constant ${\cal N}_T$ can be computed in the same
way as its analog in eq. (\ref{99}):

\bea
{\cal N}_T^{-1}&=&\left(- \frac{N}{2\pi^2T^2}\right)^{-1}
\int\limits_{q(0)=q(1)=x} Dq\:\int_{PBC}D^2\psi D^2 z\delta\left(
z^{\dag}z -1-\frac N 2 \right)\nn\\
&&\delta\left( \bar{\psi}\psi
-3\right)
\exp \left\{ i
\intt \dt \left[ -\frac{\dot q^2}{4T}+i\zd \dot z
+i\psi^{\dag}\dot{\psi}\right]\right\}
\la{114}
\eea

Substituting (\ref{113}) in (\ref{107}), we obtain desired
representation for fer\-mi\-o\-nic determinant:

{\samepage
 \bea
\lefteqn{
\ln{\det}(i\nh-m)}\nn\\
&=&\dTM\dQM\nn\\
&&{\cal N}_T \exp\left\{i
\intt\dt\left[-\frac{\dot q^2}{4T}+i\zd z
+i\bar{\psi}\dot{\psi}+\right.\right.\nn\\
&&\left.\left.\zd \dot q A(q) z+ T(\bar{\psi}\sigma^{\mu\nu} \psi)(\zd
F^{\mu\nu}(q)z)\right]\right\}
\la{115}
\eea

}

The representation (\ref{115}) is manifestly gauge and Lorentz
invariant and comprises only bosonic variables. The "action" in
(\ref{115}) is real. So, as we will see soon, it is convenient for
application of stationary phase method.

Now let us derive bosonic path integral representation for
generating functional $Z(j)$ of gauge invariant Green functions

\be
G_{f_1,\ldots,f_n}(x_1,\ldots,x_n)=
<T(\bar{\psi}_{f_1}(x_1)\psi_{f_1}(x_1)\ldots
\bar{\psi}_{f_n}(x_n)\psi_{f_n}(x_n))>
\la{116}
\ee

The derivation is completely analogous to one given in the previous
section for corresponding Euclidean correlators.

For $Z(j)$ there exist standard path integral representation via
anti-commuting variables:

\bea
Z(j)&=&\int DA e^{iS_{YM}}\int D\bar{\Psi}D\Psi
e^{
i\int dx\: \sum\limits_f \left[\bar{\Psi}_f(i\nh-m_f)\Psi_f+
j_f\bar{\Psi}_f\Psi_f\right]}\nn\\
&=&
\int DA e^{iS_{YM}}\prod_f{\det}(i\nh-m_f+j_f)
\la{117}
\eea

Repeating with minor changes the derivation of (\ref{106}), one gets

{\samepage
\bea
\lefteqn{
\ln{\det}(i\nh-m-j)}\nn\\
&=&\dTM\dQM\nn\\
&&{\cal N}_T \exp\left\{i\intt\dt\left[-\frac{\dot q^2}{4T}+i\zd z
+i\bar{\psi}\dot{\psi}+
\zd \dot q A(q) z+ \right.\right.\nn\\
&&\left.\left.T(\bar{\psi}\sigma^{\mu\nu} \psi)(\zd
F^{\mu\nu}(q)z)+
2mTj(q)-iT\bar{\psi}\g^{\mu}\psi\partial_{\mu}j(q)-
Tj^2(q)
\right]\right\}\nn\\
&&\la{118}
\eea

2}

Substituting (\ref{118}) in (\ref{117}), we obtain worldline pat
integral representation for $Z(j)$.

Our next task is the investigation of quasiclassical approximation in
QCD. To this end, we will formulate a scheme of evaluation of
two-point function

\be
G_{f_0}(x,y)=<T(\pbb_{f_0}(x)\psi_{f_0}(x)\pbb_{f_0}(y)\psi_{f_0}(y))>
\la{119}
\ee

This scheme can be easily generalized for evaluation of arbitrary
   Green functions.

The equations for the stationary point will be interpreted as
quasiclassical equations in QCD. They will be formulated in terms of
particles that have spin and colour degrees of freedom and interecting
with Yang-Mills field.

First of all, we introduce more condenced notations:

\bea
Q_f&\equiv& \{q_f,\zd_f,z_f,\pbb_f,\psi_f,T_f,m^2_f\}
\la{120.0}\\
\int DQ_f(\cdots)&\equiv&
-\frac 1 2 \int_0^{\infty} \frac{dT_f}{T_f} \:
\int_{PBC} DqD^2\psi D^2 z\delta\left( z^{\dag}z
-1-\frac N 2 \right)\nn\\
&&\delta\left( \bar{\psi}\psi -3\right)
{\cal N}_T(\cdots)
\la{122.1}\\
S[Q_f,A]&=&
\intt\dt\left[-\frac{\dot q_f^2}{4T_f}+i\zd_f\dot z_f
+i\bar{\psi_f}\dot{\psi_f}+
\zd_f \dot q A(q) z_f
\right.\nn\\
&&\left.+ T_f(\bar{\psi_f}\sigma^{\mu\nu}
\psi_f)(\zd_f F_{\mu\nu}(q)z_f)-T_fm^2_f\right]
\la{120}
\eea

Further, using eq. (\ref{118}), one can get:

\be
\frac{\dl}{\dl j(x)}
\ln\det (i\nh -m_f+j)
\raisebox{-10pt}{\rule{0.4pt}{20pt}${\scriptstyle j=0}$}
=\int DQ_f e^{iS[Q_f,A]}R(x|Q_f)
\la{121}
\ee

{\samepage
\bea
&&\frac{\dl^2}{\dl j(x)\dl j(y)}
\ln\det (i\nh -m_f+j)
\raisebox{-10pt}{\rule{0.4pt}{20pt}${\scriptstyle j=0}$}
\nn\\
&=&\int DQ_f e^{iS[Q_f,A]}R(x|Q_f)R(y|Q_f)\nn\\
&+&i\dl(x-y)\int DQ_f e^{iS[Q_f,A]}T_f\intt dt_1dt_2\:
\dl(q_{f}(t_1)-q_f(t_2))
\la{122}
\eea

}

where

\be
R(x|Q_f)= T\intt
\dt[2m_f-i\bar{\psi}_f(t)\g^{\mu}\psi_f(t)\partial_{\mu}]
\dl(x-q_f(t)) \la{123} \ee

For any functional $W(j)$

\be
\frac{\dl^2 W(j)}{\dl j(x)\dl j(y)}=
W(j)\left[
\frac{\dl^2 \ln W(j)}{\dl j(x)\dl j(y)}+
\frac{\dl \ln W(j)}{\dl j(x)}
\frac{\dl \ln W(j)}{\dl j(y)}
\right]
\la{124}
\ee

Using (\ref{117}), (\ref{121}), (\ref{122}) and applying (\ref{124})
for $W(j)=\det(i\nh-m_{f_0}+j)$, one obtains:

\be
G_{f_0}(x-y)=G^{(1)}_{f_0}(x-y)+G^{(2)}_{f_0}(x-y)
\la{125}
\ee

\noi where $G^{(1)}_{f_0}$ and $G^{(2)}_{f_0}$ correspond to the first
and to the second terms in R.H.S. of (\ref{124}) respectively:

\bea
G^{(1)}_{f_0}(x-y)&=&
\int DADQ_{f_0}\:
R(x,Q_{f_0})R(y,Q_{f_0})
\nn\\
&&\exp\left\{
iS_{YM}+iS[Q_{f_0},A]
+ \sum_f\int DQ_f \: e^{iS[Q_f,A]}\right\}\nn\\
&&+\dl(x-y)(\cdots)
\la{126}\\
G^{(2)}_{f_0}(x-y)&=&
\int DADQ_{f_0}D{Q'}_{f_0}\:
[R(x,Q_{f_0})R(y,Q_{f_0})\nn\\
&&\exp\left\{\vphantom{
\sum_f\int DQ_f \: e^{iS[Q_f,A]}}
iS_{YM}+iS[Q_{f_0},A]+iS[Q^{'}_{f_0},A]\right.\nn\\
&&\left.+ \sum_f\int DQ_f \: e^{iS[Q_f,A]}\right\}
\la{127}
\eea

\noi In (\ref{126}) $(\cdots)$ means the factor at $\dl(x-y)$ in
R.H.S. of (\ref{122}).

At first, we investigate the function $G^{(1)}_{f_0}$.

The function $G_{f_0}(x-y)$, in itself, is defined up to counterterm

$$const\, \dl(x-y)$$

\noi  by virtue of ultraviolet divergences. Then the last term in
R.H.S.  of (\ref{126}) only redefines this counterterm and so can be
omitted.

Expanding

\be
\exp\left\{\sum_f\int DQ_f\: e^{iS[Q,A]}\right\}
\la{128}
\ee

\noi in series, we can represent (\ref{126}) as

\bea
\lefteqn{
G^{(1)}_{f_0}(x-y)}\nn\\
&=&\sum_{n_f}\frac 1{\prod_f n_f!} \int DADQ_{f_0}\:
\left( \prod_f \prod_{j_f=1}^{n_f}DQ_{j_f}\right)
R(x,Q_{f_0})R(y,Q_{f_0})
\nn\\
&&\exp\left\{iS_{YM}+iS[Q_{f_0},A]+ \sum_f\sum_{j_f=1}^{n_f}
iS[Q_{j_f},A]\right\}
\nn\\
&\equiv&\sum_{n_f}G^{(1)}_{f_0}(x-y;\{n_f\})
\la{129}
\eea

Obviously, each function $G^{(1)}_{f_0}(x-y;\{n_f\})$ correspond to
    contribution of all diagrams comprising $n_{f_1}$ quark loops of
    flavor $f_1$, $n_{f_2}$ quark loops of flavor $f_2$, etc.

We will investigate each term in the series (\ref{129}) separately. At
    first we consider the term with $n_f=0$:

\be
G^{(1)}_{f_0}(x-y;\{n_f=0\})=
\int DADQ_{f_0}\: e^{iS_{YM}+iS[Q_{f_0},A]}
R(x,Q_{f_0})R(y,Q_{f_0})
\la{130}
\ee

The function $R(x|Q_f)$ can be represented as

\be
R(x|Q_f)=\int d^4p\:\intt\dt\tilde R (p,t|Q_f)e^{ip(x-q_f(t))}
\la{131}
\ee

\noi where

\be
\tilde R (p,t|Q_f)=T(2m_f+p_{\mu}\pbb(t)\g^{\mu}\psi(t))
\la{132}
\ee

\noi Then we change variables:

\be
q(t)\to q'(t)=q(t)-q_0, \ \ A(x)\to A(x-q_0)
\la{132'}
\ee

\noi where the function $q'(t)$ obeys the boundary conditions

\be
q'(0)=q'(1)=0
\la{132''}
\ee

By virtue of translational invariance we get

\bea
\lefteqn{
G^{(1)}_{f_0}(x-y;\{n_f=0\})
}\nn\\
&=&\int DA\int_{q(0)=q(1)=0}DQ_{f_0}\:\int dq_0\:
e^{iS_{YM}+iS[Q_{f_0},A]}\nn\\
&&\int d^4p_1d^4p_2 \intt dt_1\:\intt dt_2\:
\tilde R(p_1,t_1|q_{f_0}) R(p_2,t_2|q_{f_0})\nn\\
&&e^{ip_1(y-q_{f_0}(t_1))}e^{ip_2(y-q_{f_0}(t_2))}
e^{-i(p_1+p_2)q_0}\nn\\
&&
\la{133}
\eea

Integration over $q_0$ gives $\dl(p_1+p_2)$, and we obtain the
following representation for Fourier transformation of
$G^{(1)}_{f_0}(x-y;\{n_f=0\})$:

\bea
\lefteqn{
G^{(1)}_{f_0}(p;\{n_f=0\})
}\nn\\
&\equiv&\int dx\: e^{-ipx}
G^{(1)}_{f_0}(x;\{n_f=0\})
\nn\\
&=&\intt {\dt}_1 \intt{\dt}_2 \int DA\int_{q_{f_0}(0)=
q_{f_0}(1)=0} DQ_{f_0}\:
\tilde R (p,t_1|Q_{f_0})\tilde R(-p,t_2|Q_{f_0})\nn\\
&&\exp \left\{ iS_{YM}+iS[Q_{f_0},A]-ip(q_{f_0}(t_1)-q_{f_0}(t_2))
\right\}
\la{134}
\eea

Now it is already easy to write equations for stationary point for
the action in (\ref{134}). Omitting the index $f_0$, we get:

\bea
\frac 1{g^2}\nabla_{\nu}F^{a\mu\nu}&=&
\intt \dt I^a(t)\dot q^{\mu}(t)\dl(x-q(t))\nn\\
&&+T\nabla_{\nu}\left[ \intt \dt I^a(t) S^{\mu\nu}
\dl(x-q(t))\right]
\la{135}
\eea

\be
i\left(
\frac d{dt} -i\dot q^{\mu}A_{\mu}(q)
\right)z + TS^{\mu\nu}F_{\mu\nu}z=0
\la{136}
\ee

\be
i\frac d{dt}\psi +TF^a_{\mu\nu}I^a\sigma^{\mu\nu}\psi=0
\la{137}
\ee

\be
\frac 1 T \ddot q_{\mu}+\dot q^{\nu}F^a_{\mu\nu}I^a
+\nabla_{\mu}F^a_{\nu\rho}(q)I^a S^{\nu\rho}=p_{\mu}(\dl(t_1)-
\dl(t_2))
\la{138}
\ee

\be
\frac 1{4T^2} \intt \dt \dot q^2 +\frac 1 2
\intt \dt S^{\mu\nu}F^a_{\mu\nu}I^a=m^2
\la{139}
\ee

\noi where

\be
I^a=\zd \lm^a z,\ \ S^{\mu\nu}=\bar{\psi}\sigma^{\mu\nu}\psi
\la{140}
\ee

Eqs. (\ref{135})-(\ref{138}) can be derived by variation of the
action in (\ref{134}) with respect to $A^a_{\mu},\ z,\ \psi$, and
$q^{\mu}$. In the derivation of (\ref{138}) we used
(\ref{135})-(\ref{137}). The eq. (\ref{139}) is obtained by
differentiation with respect to $T$.

It easy to obtain closed system of equations in terms of
$A^a_{\mu},\ z,\ I^a,$ and $S^{\mu\nu}$.

Let

\be
I=\lm^aI^a
\la{141}
\ee

\noi Then

\be
iDI=TS^{\mu\nu}[F_{\mu\nu},I]
\la{142}
\ee

\be
\frac d{dt}S^{\mu\nu} =2TI^a F^{a[\mu}_{\ \ \ \ \rho}S^{\nu]\rho}
\la{142'}
\ee

\noi where

\be
DI\equiv \frac{dI}{dt}-i[\dot q^{\mu}A_{\mu}(q),I]
\ee

Unknown functions in (\ref{135})-(\ref{138}) also obey boundary
conditions

\be
q(0)=q(1)=0,\ \ z(0)=z(1),\ \ \psi(0)=\psi(1)
\la{144}
\ee

\noi They also satisfy the equations

\be
\zd z=1+\frac N 2,\ \ \pbb\psi=3
\la{145}
\ee

\noi because of the presence of $\dl$-functions in the
definition (\ref{122.1}) of the measure $DQ$. Apropos, we didn't
introduce Lagrange multipliers to take into account these
$\dl$-functions because the solutions of eqs. (\ref{136}),
(\ref{137}) automatically satisfy the conditions

\be
\zd z=const,\ \ \pbb \psi=const
\la{146}
\ee

Equations (\ref{135}), (\ref{138}), and (\ref{142}) are nothing but
generalization of well-known Wong's equations \cite{24} that describe
classical spinless particle interacting with Yang-Mills field.
Indeed, if one omits the terms containing the tensor of
spin $S^{\mu\nu}$ in eqs. (\ref{135}), (\ref{138}), and (\ref{142})
one gets just Wong's equations up to term

\be
-p_{\mu}(\dl(t_1)-\dl(t_2))
\la{147}
\ee

Eqs. (\ref{135})-(\ref{137}) admit simple interpretation. At "time"
$t_1$ quark--anti-quark pair with momentum $p$ is created. Then quark
and anti-quark move interacting with Yang-Mills field. The union of
quark and anti-quark trajectories forms a closed loop passing the
point $q=0$. (See (\ref{144})). At the "time" $t_2$ the
quark--anti-quark pair annihilates.

An analog of eqs. (\ref{135})-(\ref{139}) for terms with $n_f\ne 0$ in
expansion (\ref{129}) can be derived in the similar way. Let
$S^{\{n_f=0\}}$ be the action in the formula (\ref{134}) and
$J^{a\mu}[Q]$ be the current in R.H.S. of eq. (\ref{135}). Then the
analog of $S^{\{n_f=0\}}$ for the term with non zero numbers
$\{n_f\}$ in (\ref{129}) is

\be
S^{\{n_f\}}=S^{\{n_f=0\}}+\sum_f\sum_{j_f=1}^{n_f}S[Q_{j_f},A]
\la{148}
\ee

So instead of (\ref{135}) we have in this case equations

\be
\frac 1{g^2}\nabla_{\nu}F^{a\mu\nu}=J^{a\mu}[Q_{f_0}]+
\sum_f\sum_{j_f=1}^{n_f}J^{a\mu}[Q_{j_f}]
\la{150}
\ee

Equations

\be
\frac{\dl S^{\{n_f\}}}{\dl \psi_{j_f}}=0, \ \
\frac{\dl S^{\{n_f\}}}{\dl z_{j_f}}=0, \ \
\frac{\dl S^{\{n_f\}}}{\dl T_{f}}=0
\la{150'}
\ee

\noi coincide in form with (\ref{137}), (\ref{139}), and (\ref{140}),
whereas the equation

\be
\frac{\dl S^{\{n_f\}}}{\dl q_{j_f}}=0
\la{151}
\ee

\noi differs from (\ref{138}) by the term (\ref{147}).

We see that quasiclassical configurations that give the main
contribution in functional integrals (\ref{129}) are defined by
equations of very similar structure. The same statement is valid for
semiclassical configuration that give the main contribution in the
function $G^{(2)}(x-y)$ as well as in any other gauge invariant Green
functions. Therefore proposed scheme seems to be sufficiently general
for application in QCD and other gauge theories.

In this paper we restrict ourselves to formulation of general scheme
of quasiclassical approximation in QCD leaving elaborating of details
as well as applications for forthcoming papers. So at this point we
stop our investigation of quasiclassical approximation in QCD. Brief
discussion of possible applications will be given in the next
section.

\section{Conclusion}

In the present paper we derived, first, new path integral
representations for path ordered exponent (see eqs.
(\ref{35})-(\ref{37.1}), (\ref{39}), (\ref{39.4}), (\ref{58})).
We give two alternative, entirely independent derivation of
these representations. So these results seems quite reliable.

Then we applied these representations to derive new variant of
non-Abelian Stokes theorem. Our result is represented by formula
(\ref{75}). Then we transformed the latter to obtain another
formulation of non-Abelian Stokes theorem that is similar to one
proposed recently by Dyakonov and Petrov \cite{21}. As a result,
we got corrected and generalized version of the theorem proved in
     \cite{21}.

Dyakonov-Petrov version of non-Abelian Stokes theorem was already
used in discussion of the role of monopoles in QCD \cite{21} and in
attempts to derive string-like effective action in framework of QCD
\cite{25}. So one can hope that our more general and simple version
of this theorem will be also useful in discussion of various problems
of QCD.

Further, we derived pure bosonic worldline path integral
representations for fermionic determinants as well as fermionic Green
functions in Euclidean QCD. (See
 eqs. (\ref{99}), (\ref{105'}), and
(\ref{106})).

This representations comprise only integrations with respect to
bosonic variables. On a finite lattice all integrals are quite simple
and well convergent. (Remind, that domain of integration with respect
to $z$ and $\psi$ in
 eqs. (\ref{99}), (\ref{105'}), and
(\ref{106}) is bounded).
So representation derived seem quite appropriate for lattice
simulations.

Our results for fermionic determinant and Green functions can be used
also in another way.

Namely, if one substitute instead of Wilson loop some phenomenological
anzatz, one obtain formulation of the theory purely in terms of point
particles. The most well-known example of such anzatz is Wilson area
law:

\be
<\trp \exp\left\{ i\oint \dxm A_{\mu}\right\}>=
\exp\left\{ -KS_{min}+\left[{\mbox{perturbative}\atop
\mbox{corrections}}\right]\right\}
\la{152}
\ee

This anzatz was applied, in particular, to derivation from QCD a
quark--anti- quark potential used in potential models (see, for
instance, recent papers \cite{26}, \cite{26.1} and references
therein).

Other anzatz for Wilson loops were proposed in framework of Dual QCD
model \cite{27} and stochastic vacuum model \cite{28}. (See also
\cite{26.1} for comparison of results obtained in framework of these
models). Recently a sting-like expression in spirit of stochastic
vacuum model was derived in papers \cite{25}, \cite{29}.Anzatz of
another type, that gives an expression for Wilson loop in terms of
trajectories of monopoles, was proposed in \cite{21}.

All these results could be combined with ours to derive some
effective action in terms of point particles that correctly describe
colour and spin properties of quarks. Obviously, that such theory is
much more simpler than initial quantum field theory and so thus
approach of investigation seems to be rather perspective.

Finally, in the present paper we also derived bosonic worldline path
integral representation for fermionic determinants and Green functions
in Minkowski space and started the investigation of the
quasiclassical approximation in QCD.

The key point in the formulation of quasiclassical approximation is
quasiclassical QCD equations (\ref{135})-(\ref{139}) and
(\ref{148})-(\ref{151}) which arise naturally when one applies the
stationary point method to evaluation of functional integrals that
defines Green functions in QCD. Quasiclassical equations derived
appear to be nothing but a generalization of well-known Wong's
equations.

We formulated only those quasiclassical equations which arise in the
problem of evaluation of the concrete Green function (\ref{119}).
However, our method is quite general and one can easy to derive
analogous equations in generic case.

The next problem in investigation of quasiclassical approximation in
QCD is the solution of quasiclassical equation of motion. Though
these equations are very complicated, this problem doesn't seem
hopeless, at least, in non-re\-la\-ti\-vis\-tic approximation. In
electrodynamics in non-re\-la\-ti\-vis\-tic approximation one can
neglect bremsstrahlung and retarded effects and, as a result,
reformulate initial theory in terms of particles interacting by means
of Coulomb forces. In the same way one may hope to derive potential
of interaction of heavy quarks in QCD. Such approach is alternative
to ones based on various anzatz for Wilson loops.

Another interesting possibility to understand the quark confinement
in framework of quasiclassical approach is connected with existence of
classical solutions of Yang-Mills equations with singularity on the
sphere. Such solution were discussed in the context of the problem of
confinement recently in the papers \cite{29'} though the existence of
such solutions was pointed out by several authors in 70's \cite{30}.
Some solutions with singularity on the torus and cylinder was
discussed in papers \cite{31}.

If solutions with singularity on closed spacelike surface
existed also  for eqs. (\ref{135})-(\ref{139}) and their
modifications, then quarks, moving inside such surface and
interacting with Yang-Mills field, couldn't cross the surface. This
would mean the confinement of quarks. So it is interesting to
investigate singular solutions of equations  (\ref{135})-(\ref{139})
and, if such solutions exist, to develop the corresponding
quasiclassical perturbative theory.

\section*{Appendix A}

In this appendix we shall prove the validity of the identity
(\ref{55}).

\noi We will prove (\ref{55}) by induction. For $N=2$ it is easy to
prove (\ref{55}) by direct calculations because both sides of
(\ref{55}) are simple rational functions of variables $\exp\{\frac i
2 \al_r\}$.

Let us denote

$$
x_r=e^{i\al_r}
\eqno(A1)
$$

\noi Then L.H.S. of (\ref{55}) can be represented as

$$
(-2i)^{N-1}
\frac{
\prod\limits_{r=1}^N \sqrt x_r}{\prod\limits_{1\le p<q\le N}(x_p-x_q)}
\sum_{j=1}^N(-1)^{j+1}x_j^N
 \prod_{{\scriptstyle {r,s\ne j\atop
1\le r<s\le N}}}(x_r-x_s)
\eqno(A2)
$$

\noi whereas R.H.S. of (\ref{55}) as

$$
(-2i)^{N-1}\left( \prod_{k=1}^N\sqrt x_k\right)
\sum_{j=1}^N x_j
\eqno(A3)
$$

\noi So eq. (\ref{55}) is equivalent to algebraic identity

$$
\sum_{j=1}^N(-1)^{j+1}x_j^N \prod_{{\scriptstyle {r,s\ne j\atop
1\le r<s\le N}}}(x_r-x_s)
=
\left(\sum_{j=1}^N x_j\right)
\prod\limits_{1\le r<s\le N}(x_r-x_s)
\eqno(A4)
$$

To prove (A4), we check, at first, that L.H.S. of (A4) is vanished if
$x_i=x_j$. Obviously, it is sufficient to consider the case
$x_1=x_2\equiv x$.

If $x_1=x_2$ then only $j=1$ and $j=2$ terms survive in L.H.S. of
(A4). The $j=1$ term is

$$
x_1^N\prod\limits_{2\le r<s\le N}(x_r-x_s) =
x^N\left(\prod_{p=3}^N(x-x_p)
\right)\left(\prod\limits_{3\le p<q\le N}(x_p-x_q)\right)
\eqno(A5)
$$

\noi whereas the $j=2$ term is equal to

$$
-x_2^N
 \prod_{{\scriptstyle {r,s\ne 2\atop
1\le r<s\le N}}}(x_r-x_s)
=-
x^N\left(\prod_{p=3}^N(x-x_p)
\right)\left(\prod\limits_{1\le p<q\le N}(x_p-x_q)\right)
\eqno(A6)
$$

\noi So terms with $j=1$ and $j=2$ are cancelled.

Thus L.H.S. of (A4) can be represented as

$$
P(x_1,\ldots,x_N)
\prod\limits_{1\le r<s\le N}(x_r-x_s)
\eqno(A7)
$$

\noi We must prove that

$$
P(x_1,\ldots,x_N)
=\sum_{r=1}^N x_r
\eqno(A8)
$$

L.H.S. of eq. (A4) is polynomial of degree $N$ with respect to each
variable $x_j$. Consequently, polynomial
$P(x_1,\ldots,x_N)$
                   is linear in each $x_j$. So

$$
P(x_1,\ldots,x_N)
=a(x_2,\ldots,x_N)x_1+b(x_2,\ldots,x_n)
\eqno(A9)
$$

Comparing coefficient at $x_1^N$ in L.H.S. of (A4) and (A7), one find
that

$$a(x,\ldots,x_N)=1
\eqno(A10)
$$

\noi and so it remains to prove that

$$b(x_2,\ldots,x_N)=\sum_{j=2}^N x_j
\eqno(A11)
$$

To prove (A11), let us compare the sums of $x_1$-independent terms in
L.H.S. of (A4) and in (A7). They can be written as

$$
\sum_{j=2}^N (-1)^{j+1} x_j^N
\prod\limits_{{\scriptstyle {2\le r\le N \atop r\ne j}}}
(-x_r)
\prod\limits_{{\scriptstyle {2\le r<s\le N \atop r,s\ne j}}}
(x_r-x_s)
\eqno(A12)$$

\noi and

$$b(x_2,\ldots,x_N)\prod\limits_{2\le r\le N}
(-x_r)\prod_{2\le r<s\le N}(x_r-x_s)
\eqno(A13)
$$

\noi respectively.

So (A11) is equivalent to

$$
\sum_{j=2}^N(-1)^{j}x_j^{N-1} \prod_{{\scriptstyle {r,s\ne j\atop
2\le r<s\le N}}}(x_r-x_s)
=
\left(\sum_{j=2}^N x_j\right)
\prod\limits_{2\le r<s\le N}(x_r-x_s)
$$

But the latter equation is valid by virtue of induction assumption.
Indeed, it can be obtained from (A4) by change $N\to N-1,\ ,x_1\to
x_2, \ldots,x_{N-1}\to x_N$. This finishes the proof of identity
(\ref{55}).

\section*{Appendix B}

In this appendix we prove the formula (\ref{66}). First, one notes
that up to inessential factor

$$\det\left( -\frac d{dt} + iBe^{-\e\frac d{dt}}\right)
=\det
\oeo
\eqno(B1)
$$

\noi So it is sufficient to prove that

$$\lime \ln\det \oeo-\ln\det\owe=
\frac i 2 \intt\dt\tr B
\eqno(B2)$$

\noi or, equivalently,

$$
\lime \tr\ln\oeo-\tr\ln\owe=\frac i 2 \intt\dt\tr B
\eqno(B3)
$$

Let $\hat{\Pi}$ is orthogonal projector in $L^2(S^1)$ on any
{\it finite} dimensional subspace. Then

$$\lime \left[\tr \hat{\Pi} \ln\oeo -\tr\hat{\Pi}\ln\owe
\right]=0
\eqno(B4)$$

\noi In particular, let ${\cal H}_1$ be subspace in $L^2(S^1)$
that is orthogonal to one dimensional subspace spanned on the
function $\phi_0(t)\equiv 1$. Then

$$\lime \left[\tr  \ln\oeo -\tr\ln\owe
\right]$$
$$=\lime \left[{\tr}_{{\cal H}_1}  \ln\oeo -{\tr}_{{\cal H}_1}\ln\owe
\right]
\eqno(B5)$$

In the space ${\cal H}_1$ the operator

$$-\frac d{dt} e^{\e\frac d{dt}}
\eqno(B6)$$

\noi is invertible. Let $G_{\e}(t-t')$ is the kernel of

$$\left(-\frac d{dt} e^{\e\frac d{dt}}\right)^{-1}
$$

The function  $G_{\e}(t-t')$ can be expresseed in terms of
eigenfunctions

$$\phi_n(t)=e^{2\pi int}
,\ \ n\ne 0
\eqno(B7)$$

\noi of the operator (B6) and its eigenvalues

$$\lm_n=-2\pi in e^{2\pi in\e},\ \ n\ne 0$$

Obviously,

$$G_{\e}(t-t')=\sum_{n\ne0}\frac{\phi_n(t)\bar{\phi_n}(t')}
{\lm_n}
=-\frac 1{\pi}\sum_{n=1}^{\infty}
\frac{\sin 2\pi n(t-t'-\e)}{n}
\eqno(B8)$$

Using the formula

$$\sum_{n=1}^{\infty}\frac{\sin 2\pi n\e}{\pi n}=
\frac 1 2 -\e
\eqno(B9)$$

\noi (that is valid for $0<\e<1$), one finds that

$$\lime G_{\e}(0)=\frac 1 2
\eqno(B10)$$

\noi whereas

$$G_{0}(0)=0
\eqno(B10')$$

\noi So

$$G_{0}(0)\ne\lime G_{\e}(0)
\eqno(B11)
$$

But for $-1<t<1,\ \ t\ne0$,

$$\lime G_{\e}(t)=G_{0}(t)
\eqno(B11')$$

We will see soon that R.H.S. of (B2) is not vanished just by virtue
of (B11).

Further, up to inessential constant,

$${\tr}_{{\cal H}_1}\ln\oeo={\tr}_{{\cal H}_1}\ln\left[
1+i
\left(-\frac d{dt} e^{\e\frac d{dt}}\right)^{-1}
B\right]$$

$$={\tr}_{{\cal H}_1}\left[
\left(-\frac d{dt} e^{\e\frac d{dt}}\right)^{-1}
B\right]+
\frac 1 2 {\tr}_{{\cal H}_1}
\left(-\frac d{dt} e^{\e\frac d{dt}}\right)^{-1}
B
\left(-\frac d{dt} e^{\e\frac d{dt}}\right)^{-1}
+\ldots
\eqno(B12)$$

By virtue of equation

$$\intt \dt G_{\e}(t-t')=0
\eqno(B13)$$

\noi one can replace ${\tr}_{{\cal H}_1}$ by $\tr$ in all terms in the
series (B12). So

$${\tr}_{{\cal H}_1}\ln \oeo
=
i\intt \dt G_{\e}(0) \tr B(t)$$

$$+\frac 1 2 \intt\dt\intt dt'\: G_{\e}(t-t')
G_{\e}(t'-t)\tr B(t)B(t')+\ldots
\eqno(B14)$$

In the limit $\e \to +0$ the first term in R.H.S. of (B14) is equal
to

$$\frac i 2 \intt\dt \tr B(t)$$

\noi (see (B11)) whereas the sum of all others coincide with

$${\tr}_{{\cal H}_1}\owe$$

\noi by virtue of (B10'), (B11'). This proves the validity of the
equation (B2).

To check the formula (B2), let us consider the simplest case of one
dimensional harmonic  oscillator. Partion function

$$Z(\beta)=\tr\exp\{-\beta a^{\dag} a\}
\eqno(B15)$$

\noi is known exactly:

$$Z(\beta)=\sum_{n=0}^{\infty}e^{-\beta n}=\frac 1{1-e^{-\beta}}
\eqno(B15)$$

On the other hand,

$$Z(\beta)=\lime\int_{PBC}D\bar z Dz\:
\exp\left\{\intt\dt(-\bar z\dot z -\beta \bar z e^{-\e\frac d{dt}}z
\right\}$$

$$={\det}^{-1}\left(-\frac d{dt}-\beta e^{-\e\frac d{dt}}
\right)
\eqno(B16)
$$

Putting in (B2) $B=i\beta$, one gets:

$${\det}^{-1}\left(-\frac d{dt}-\beta e^{-\e\frac d{dt}}
\right)
=e^{\frac{\beta}{2}}
{\det}^{-1}\left(-\frac d{dt}-\beta
\right)
\eqno(B17)$$

The latter determinant we have already evaluated in the main text of
the paper :

$$
{\det}\left(-\frac d{dt}-\beta
\right)
=const \,\sinh \frac{\beta}2
\eqno(B18)$$

\noi (see eq. (\ref{50}) for $N=1,\ \ \eta=\e=0$).

Comparing (B16)-(B19), we get

$$Z(\beta)=const\,\frac{e^{\frac{\beta}2}}{\sinh\frac{\beta}2}
\eqno(B19)$$

Evaluating the constant from the condition

$$Z(\infty)=1$$

\noi we reproduce the true answer (B15). We would like to stress that
if we had put $\e=0$ in (B16) before evaluating of the functional
integral the result would have been wrong.

 \end{document}